\documentclass[genTeX]{nrc1}

\journal{Can. J. Phys.}
\journalcode{cjp}

\usepackage{graphicx}
\usepackage[figuresright]{rotating}
\usepackage{amsmath}
\usepackage{amssymb}
\usepackage{bm}
\usepackage{cite}

\usepackage{epsfig}
\usepackage{amsbsy}
\usepackage{latexsym}
\usepackage{amsfonts}
\usepackage{amsthm}
\usepackage{bm}
\usepackage{cite}
\usepackage[french, english]{babel}

\begin{document}

\title{Entangling identical bosons in optical tweezers via exchange interaction}
\author[N. S. Babcock]{Nathan S. Babcock}
\address[G1]{Institute for Quantum Information Science, University of Calgary, Alberta, Canada.\\ \email{nbabcock@qis.ucalgary.ca}}
\author[R. Stock]{Ren\'{e} Stock}
\address[G2]{Department of Physics, University of Toronto, Toronto, Ontario, Canada. \\\email{restock@physics.utoronto.ca}}
\author[M. G. Raizen]{Mark G. Raizen}
\address[G3]{Center for Nonlinear Dynamics and Department of Physics, University of Texas, Austin, Texas, U.S.A.}
\author[B. C. Sanders]{Barry C. Sanders}
\address[G4]{Institute for Quantum Information Science, University of Calgary, Alberta, Canada.}

\shortauthor{Babcock, Stock, Raizen, and Sanders}

\maketitle

\begin{abstract}
We first devise a scheme to perform a universal entangling gate via controlled collisions between pairs of atomic qubits trapped with optical tweezers. Second, we present a modification to this scheme to allow the preparation of atomic Bell pairs via selective excitation, suitable for quantum information processing applications that do not require universality. Both these schemes are enabled by the inherent symmetries of identical composite particles, as originally proposed by Hayes \textit{et al.} Our scheme provides a technique for producing weighted graph states, entangled resources for quantum communication, and a promising approach to performing a ``loophole free'' Bell test in a single laboratory.
\end{abstract}

\begin{resume}
   \traduit
\end{resume}

\def\tablefootnote#1{%
\hbox to \textwidth{\hss\vbox{\hsize\captionwidth\footnotesize#1}\hss}}

\section{Introduction}
Entanglement plays an indispensable role in many quantum information processing tasks, such as long-distance quantum communication~\cite{Briegel:1998}, teleportation-based quantum computation \cite{Gottesman:1999, KLM:2001}, and one-way quantum computation~\cite{Raussendorf:2001}. While great progress has been made entangling arrays of neutral atoms in optical lattices \textit{en masse} \cite{Mandel:2003}, the current approach to generating such massive entangled states (via cold collisions) necessitates state-dependent traps~\cite{Jaksch:1999}. This state-dependency results in increased noise sensitivity and decoherence of atomic qubits~\cite{Mandel:2003}. Other proposed approaches for entangling neutral atoms feature encodings in vibrational rather than internal electronic states of atoms~\cite{Charron:2002,Eckert:2002}, but are subject to similar dephasing of qubits. Approaches based on atomic interactions other than ground state collisions have been suggested~\cite{Brennen:2002,Jaksch:2000}, but none have been successfully implemented and atomic collisions still hold the most promise. Thus, there is a need for collisional quantum gates that allow more flexible encodings in robust electronic states---such as the clock states of Rb, Cs, or Group II atoms---that are held in state-insensitive traps to minimize decoherence.

In this work, we examine schemes to entangle pairs of bosonic atoms, analogous to the recently proposed fermionic spin-exchange gate \cite{Hayes:2007}. Gates based on this exchange interaction offer a natural resistance to errors and more flexibility due to inherent symmetrization conditions.  Furthermore, this exchange interaction allows the design of entangling operations for atoms with state-independent (e.g., Rb~\cite{Mandel:2003}) or partially unknown interaction strengths (e.g., Yb~\cite{Kitagawa:2007} or Sr~\cite{Yasuda:2006}). The underlying exchange interaction for these gates has recently been experimentally demonstrated using bosonic Rb atoms in a double-well optical lattice~\cite{Anderlini:2007}. However, a verifiable entangling gate between an individual pair of trapped neutral atoms has not yet been demonstrated. Here, we provide a detailed analysis of these operations as they may be carried out using a pair of individually controlled atomic qubits trapped via optical tweezers.

Our approach builds on disparate proposals and experiments for preparing individual atoms from a Bose-Einstein condensate \cite{Dudarev:2007, Chuu:2005, Diener:2002}, encoding qubits into long-lived electronic states, coherently manipulating and transporting atoms using optical tweezers \cite{Gustavson:2007,Beugnon:2007}, and performing two-qubit operations on pairs of atoms via collisional interactions \cite{Hayes:2007, Anderlini:2007}. The combination of these elements allow for the design of a tunable two-qubit gate, which can create an arbitrary degree of entanglement between a pair of atoms. We also examine a scheme that exploits symmetrization rules to produce Bell pairs via selective excitation.

These entangling schemes may be realized using qubits stored in the electronic states of a pair of atoms trapped with moveable optical tweezers. Trapping at a ``magic wavelength'' makes the light shift potential state-independent. Encoding in atomic clock states---which are insensitive to fluctuations in the trapping field---avoids dephasing and ensures qubit coherence during the transport process. Unlike the case of a state-dependent optical lattice in which it is trivial to separate the atoms after interaction, we have state-independent potentials in which the system's dynamics determine the likelihood of the atoms being separated into opposite wells. Under adiabatic conditions, atom separation is guaranteed. We consider only the 1-D case for simplicity. Multidimensional effects such as trap-induced resonances cannot be captured by the 1-D delta-potential employed here \cite{Stock:2003} but could potentially be used to enhance the atomic interaction further.

\section{Hamiltonian for identical particles in separated tweezers}

The Hamiltonian for two atoms with internal structure in a pair of optical dipole traps (a.k.a., ``tweezers'') is given by,
\begin{align}\label{hamiltonian}
H =\sum_{i,j=0,1}\bigg\{ \frac{p_\text{a}^2}{2m} + V(x_\text{a},d ) + \frac{p_\text{b}^2}{2m} + V(x_\text{b},d)
+ 2a_{ij}\hbar\omega_\bot\delta(x_\text{a}-x_\text{b}) \bigg\}  \otimes|ij\rangle\!\langle ij|,
\end{align}
where $x_\text{a}$ and $x_\text{b}$ are the positions of atoms $\text{a}$ and $\text{b}$ respectively, $p_\text{a}$ and $p_\text{b}$ are similarly the momenta, $a_{ij}$ is the state-dependent scattering length that depends on internal atomic states $|i\rangle_\text{a}$ and $|j\rangle_\text{b}$ (using $|ij\rangle \equiv |i\rangle_\text{a}\!\otimes\!|j\rangle_\text{b}$), $\omega_\bot$ is the harmonic oscillation frequency due to transverse confinement \cite{Calarco:2000}, and $d$ is the time-dependent centre-to-centre distance between wells. For Yb, Sr, and alkali atoms, one can usually choose a particular trap-laser wavelength (the ``magic wavelength'') so that the light shift potential becomes state-independent and each atom sees a double-well potential:
\begin{equation}
V(x,d) = -V_o e^{-(x-\frac{d}{2})^2/2\sigma^2} - V_o e^{-(x+\frac{d}{2})^2/2\sigma^2}.
\end{equation}
Here, $V_o > 0$ is the depth of each Gaussian well and $\sigma^2$ is the variance.

\begin{figure}[h]
  \begin{center}
  \includegraphics[width=140mm]{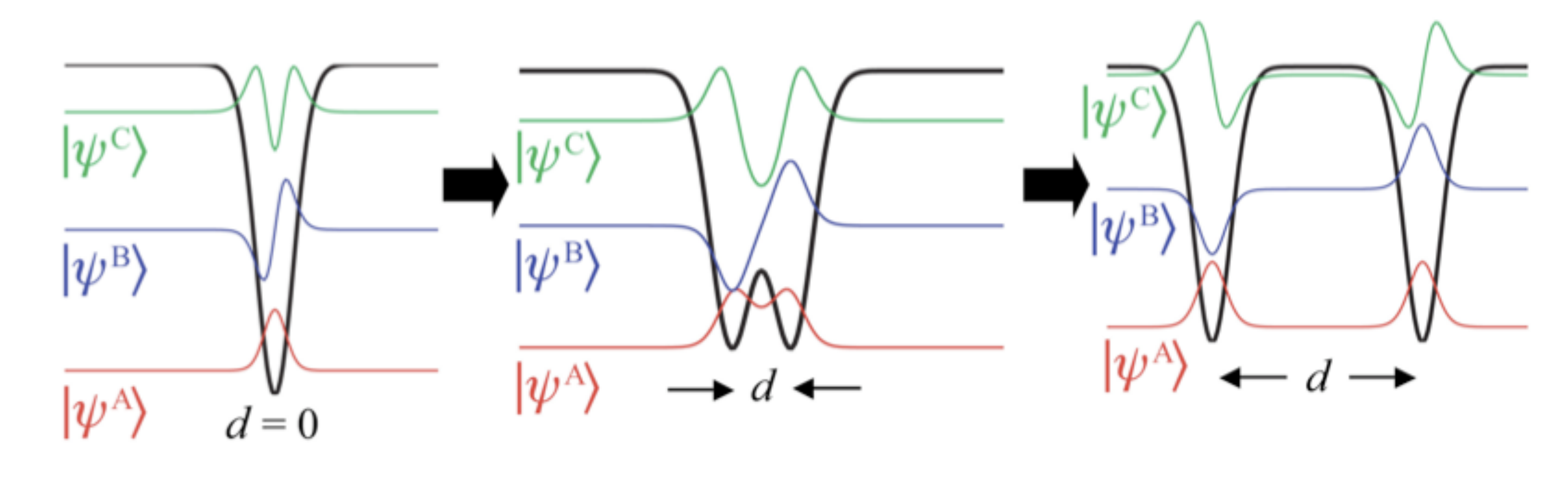}
  \end{center}
  \addtolength{\abovecaptionskip}{-5mm}
  \caption{The first three eigenstates of a single particle in a double-well potential for different well separations $d$.\label{fig1}}

\end{figure}

The first three vibrational eigenstates of a single particle in this double-well potential are shown in Fig. \ref{fig1} for varying $d$. In general, the single-particle eigenstates are $\{|\psi^A(d)\rangle, |\psi^B(d)\rangle, |\psi^C(d)\rangle,\ldots\}$ and $d$-dependence is assumed implicit (e.g., $|\psi^A\rangle \equiv |\psi^A(d)\rangle$) for notational simplicity. Note that as $d$ increases, $|\psi^A\rangle$ and $|\psi^B\rangle$ become spatially delocalized and energetically degenerate. Thus, when $|d| \gg \sigma$ we can write $|\psi^L\rangle \equiv (|\psi^A\rangle-|\psi^B\rangle)/\sqrt{2}$ to represent a single particle localized in the ground state of the left well, and similarly $|\psi^R\rangle \equiv (|\psi^A\rangle+|\psi^B\rangle)/\sqrt{2}$ for the right well.

\begin{figure}[h]
\begin{center}
$\begin{array}{c@{\hspace{1mm}}c}
\multicolumn{1}{l}{\mbox{\bf (a)}} & \multicolumn{1}{l}{\mbox{\bf (b)}} \\
    \includegraphics[width=70mm]{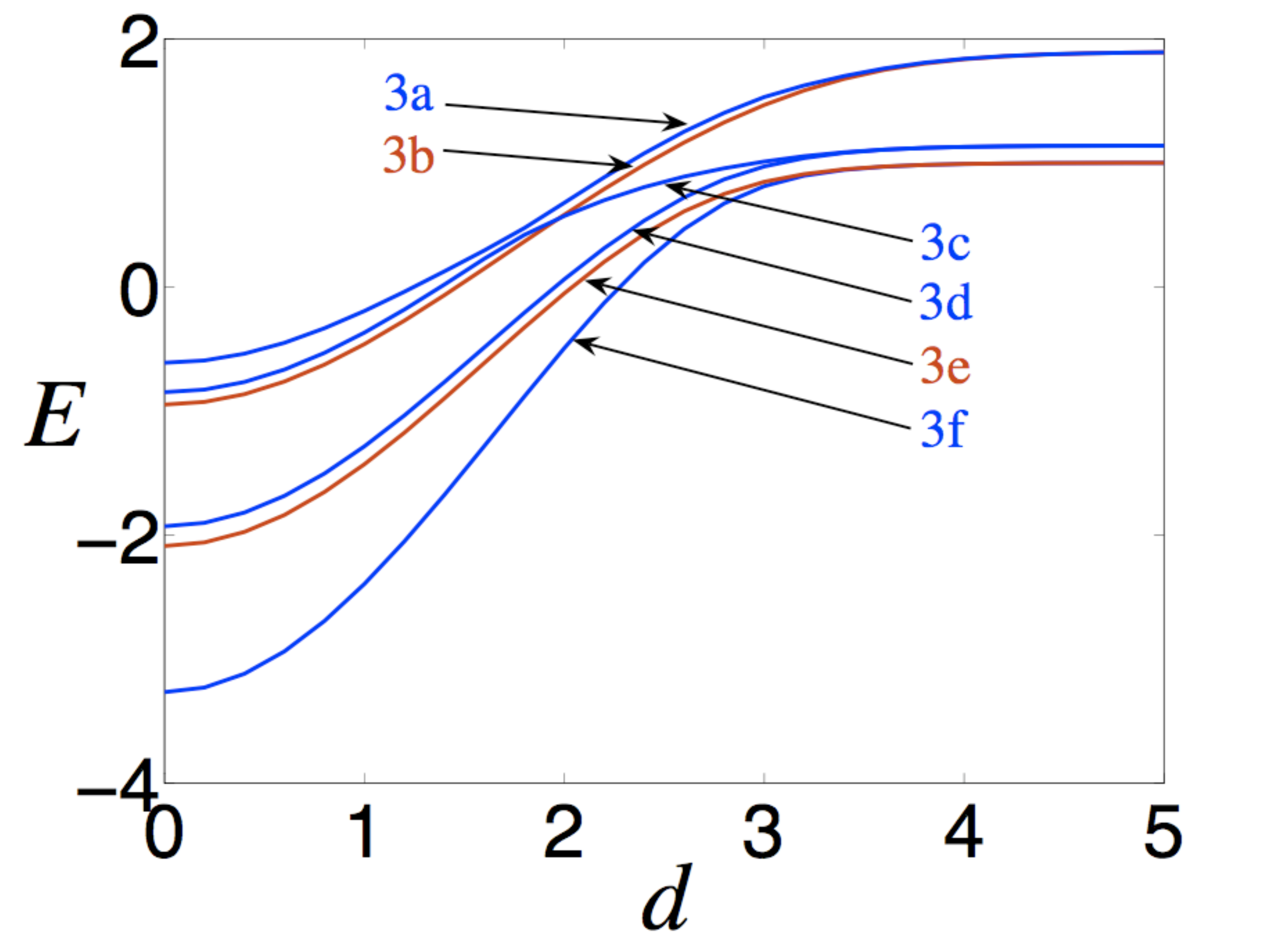} &
    \includegraphics[width=70mm]{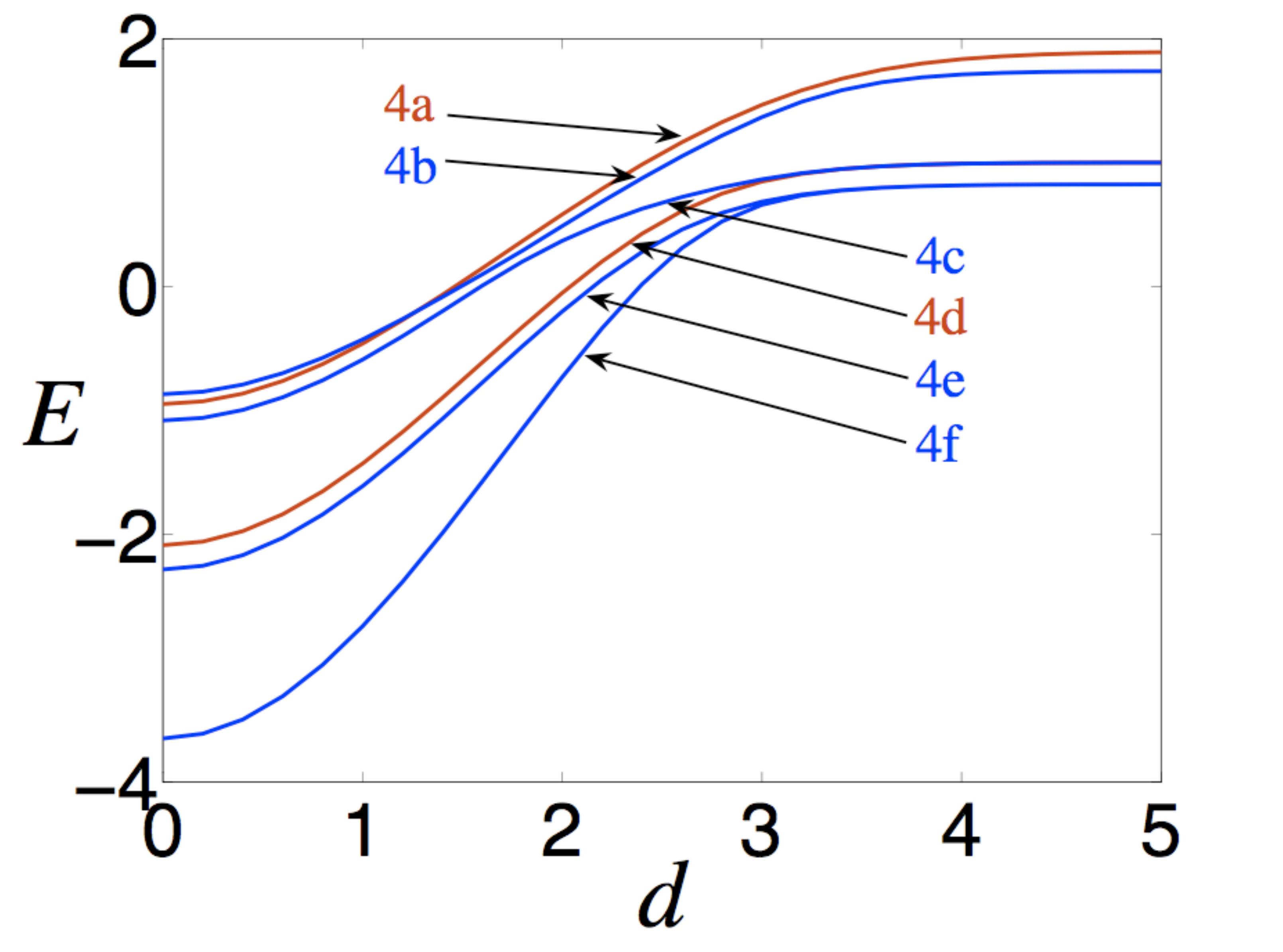}
\end{array}$
\end{center}
  \caption{Adiabatic energy levels as a function of well separation $d$ for (a) $a_{ij}=0.1\sigma$ and (b) $a_{ij}=-0.1\sigma$. Well separation is in units of $\sigma$. Energies are in units of $\hbar\omega_o$, where $\omega_o$ is the harmonic oscillation frequency of one atom in the ground state of a single well. Symmetric vibrational eigenstates are shown in blue, antisymmetric in red. Eigenstates for (a) and (b) are given by Eqs. (\ref{eigenstates0})   and (\ref{eigenstates5}), respectively. Notice that crossings between oppositely symmetrized states are unavoided because the Hamiltonian is symmetric.}
  \label{fig2}
\end{figure}

When a second particle is added to the double-well potential, the interaction term in the Hamiltonian may be treated as a perturbation. Accordingly, the new two-particle eigenstates may be written as a sum of perturbed tensor products of one-particle states. We will use a tilde to denote the perturbation to the terms composing the new symmetrized eigenstates. For a repulsive interaction between atoms ($a_{i\!j}>0$), the first six two-particle eigenstates are (see Fig. \ref{fig2}a),
\vspace{1mm}
\begin{subequations}\label{eigenstates0}
\begin{align}
d & =0 & \quad&\quad & d \; & \!\gg \sigma \notag \\
 & \!\!\!\!\!\! |\widetilde{\psi^B \psi^B}\rangle & \quad&\longleftrightarrow\quad & \textstyle\frac{1}{2}(|\widetilde{\psi^A\psi^C}\rangle + |\widetilde{\psi^C\psi^A}\rangle &- |\widetilde{\psi^B\psi^D}\rangle - |\widetilde{\psi^D\psi^B}\rangle) \\
\textstyle\frac{1}{\sqrt{2}}(|\psi^A\psi^C\rangle &- |\psi^C\psi^A\rangle) & \quad&\longleftrightarrow\quad & \textstyle\frac{1}{\sqrt{2}}(|\psi^A\psi^C\rangle &- |\psi^C\psi^A\rangle) \\
\textstyle\frac{1}{\sqrt{2}}(|\widetilde{\psi^A\psi^C}\rangle & + |\widetilde{\psi^C\psi^A}\rangle) & \quad&\longleftrightarrow\quad & \textstyle\frac{1}{\sqrt{2}}(|\widetilde{\psi^L\psi^L}\rangle &+ |\widetilde{\psi^R\psi^R}\rangle) \\
\textstyle\frac{1}{\sqrt{2}}(|\widetilde{\psi^A\psi^B}\rangle &+ |\widetilde{\psi^B\psi^A}\rangle) & \quad&\longleftrightarrow\quad & \textstyle\frac{1}{\sqrt{2}}(|\widetilde{\psi^L\psi^L}\rangle &- |\widetilde{\psi^R\psi^R}\rangle) \\
\textstyle\frac{1}{\sqrt{2}}(|\psi^A\psi^B\rangle &- |\psi^B\psi^A\rangle) & \quad&\longleftrightarrow\quad & \textstyle\frac{1}{\sqrt{2}}(|\psi^L\psi^R\rangle &- |\psi^R\psi^L\rangle) \\
& \!\!\!\!\!\! |\widetilde{\psi^A\psi^A}\rangle & \quad&\longleftrightarrow\quad & \textstyle\frac{1}{\sqrt{2}}(|\widetilde{\psi^L\psi^R}\rangle &+ |\widetilde{\psi^R\psi^L}\rangle).
\end{align}
\end{subequations}

States that are antisymmetric under exchange are not affected by the interaction at any separation and the tildes have been intentionally omitted from these states. States with atoms in opposite traps (e.g., 3a, 3f) are obviously not affected by the interaction in the limit $d\rightarrow\infty$. Note that there is the usual on-site interaction penalty for putting two atoms in same trap, resulting in an energy splitting at $d \gg \sigma$ between states having atoms in opposite traps (3e, 3f) and those having atoms in the same trap (3c, 3d), as shown in Fig. \ref{fig2}a.

In case of attractive interaction ($a_{i\!j}<0$), the eigenstates are (see Fig. \ref{fig2}b),
\begin{subequations}\label{eigenstates5}
\vspace{1mm}
\begin{align}
d & =0 & \quad&\quad & d \; & \!\gg \sigma \notag \\
\textstyle\frac{1}{\sqrt{2}}(|\psi^A\psi^C\rangle &- |\psi^C\psi^A\rangle) & \quad&\longleftrightarrow\quad & \textstyle\frac{1}{\sqrt{2}}(|\psi^A\psi^C\rangle &- |\psi^C\psi^A\rangle) \\
 & \!\!\!\!\!\! |\widetilde{\psi^B \psi^B}\rangle & \quad&\longleftrightarrow\quad & \textstyle\frac{1}{2}(|\widetilde{\psi^A\psi^C}\rangle + |\widetilde{\psi^C\psi^A}\rangle &+ |\widetilde{\psi^B\psi^D}\rangle + |\widetilde{\psi^D\psi^B}\rangle) \\
\textstyle\frac{1}{\sqrt{2}}(|\widetilde{\psi^A\psi^C}\rangle & + |\widetilde{\psi^C\psi^A}\rangle) & \quad&\longleftrightarrow\quad & \textstyle\frac{1}{\sqrt{2}}(|\widetilde{\psi^L\psi^R}\rangle &+ |\widetilde{\psi^R\psi^L}\rangle) \\
\textstyle\frac{1}{\sqrt{2}}(|\psi^A\psi^B\rangle &- |\psi^B\psi^A\rangle) & \quad&\longleftrightarrow\quad & \textstyle\frac{1}{\sqrt{2}}(|\psi^L\psi^R\rangle &- |\psi^R\psi^L\rangle) \\
\textstyle\frac{1}{\sqrt{2}}(|\widetilde{\psi^A\psi^B}\rangle &+ |\widetilde{\psi^B\psi^A}\rangle) & \quad&\longleftrightarrow\quad & \textstyle\frac{1}{\sqrt{2}}(|\widetilde{\psi^L\psi^L}\rangle &- |\widetilde{\psi^R\psi^R}\rangle) \\
& \!\!\!\!\!\! |\widetilde{\psi^A\psi^A}\rangle & \quad&\longleftrightarrow\quad & \textstyle\frac{1}{\sqrt{2}}(|\widetilde{\psi^L\psi^L}\rangle &+ |\widetilde{\psi^R\psi^R}\rangle).
\end{align}
\end{subequations}

The order of states in (\ref{eigenstates5}) is different from (\ref{eigenstates0}), since states with atoms in the same trap now posses a lower energy due to the attractive interaction.

Until this point we have neglected the internal structure of the particles (i.e., the qubits). The eigenstates of the full Hamiltonian (\ref{hamiltonian}) are tensor products of the vibrational wavefunctions and the symmetrized qubit states. For bosonic atoms, permissible eigenstates are tensor products of external (i.e., vibrational) and internal (i.e., qubit) states of the same symmetry. Thus, antisymmetrized spatial wavefunctions are permitted for a pair of composite bosons, so long as their internal structure is also antisymmetric.

We solve the Hamitonian~(\ref{hamiltonian}) numerically for individual internal states. Examples of two-atom energy spectra as a function of well separation are plotted in Fig.~\ref{fig2} for positive and negative interaction strengths $a_{ij}$. We define $E^{\,n}_{|\varphi\rangle}(d)$ to be the energy of the $n^\text{th}$ two-atom vibrational eigenstate with two-qubit internal state $|\varphi\rangle$ at well separation $d$. For example, the energy of $|\widetilde{\psi^A\psi^A}\rangle\!\otimes\!|11\rangle$ is $E^{\,0}_{|11\rangle}(d)$. As we have already discussed, not all combinations of $n$ and $|\varphi\rangle$ are possible. For example, $E^{\,0}_{|\Psi^-\rangle}(d)$ is forbidden for identical bosons. This reduction of the size of the Hilbert space makes it possible to perform quantum gates adiabatically without losing coherence due to energetic degeneracies. Furthermore, the Hamiltonian's inherent particle and parity symmetries lead to selection rules which further enhance the fidelity of two-qubit operations.

\section{Universal entangling gate}

As shown by Hayes \textit{et al.}~\cite{Hayes:2007}, it is possible to exploit these symmetries in order to produce a two-qubit entangling operation. We begin with a pair of identical atoms localized to opposite wells of the double-well Hamiltonian (\ref{hamiltonian}) in the far separated case ($d\gg\sigma$). We prepare the qubit on the left in the state $|\varphi^{\alpha}\rangle\equiv(\alpha|0\rangle+ \beta|1\rangle)$ and the qubit on the right in the state $|\varphi^{\mu}\rangle\equiv(\mu|0\rangle+ \nu|1\rangle)$. The initial wavefunction $|\psi_\text{i}\rangle$, written as a tensor product of external and internal states, is then,
\begin{gather}\label{initstate}
\left|\psi_\text{i}\right> = \textstyle\frac{1}{\sqrt{2}}(|\widetilde{\psi^L\psi^R}\rangle\otimes|\varphi^{\alpha}\varphi^{\mu}\rangle + |\widetilde{\psi^R\psi^L}\rangle\otimes|\varphi^{\mu}\varphi^{\alpha}\rangle).
\end{gather}
Using $\left|\Psi^{\pm}\right>=\frac{1}{\sqrt{2}}(\left|01\right>\pm\left|10\right>)$, we rewrite equation (\ref{initstate}) to make the symmetrization explicit:
\begin{align}
\left|\right. \left.\!\!\psi_\text{i} \right>  = (|\psi^L\psi^R\rangle-|\psi^R & \psi^L\rangle)\otimes(\textstyle\frac{\alpha\nu-\beta\mu}{2}\left|\Psi^-\right>)\;\; \\ & + (|\widetilde{\psi^L\psi^R}\rangle\!+\!|\widetilde{\psi^R\psi^L}\rangle)\!\otimes\!(\textstyle\frac{\alpha\mu}{\sqrt{2}}\!\left|00\right> \!+\! \textstyle\frac{\alpha\nu+\beta\mu}{2}\!\left|\Psi^+\right> \!+\! \textstyle\frac{\beta\nu}{\sqrt{2}}\!\left|11\right>). \notag
\end{align}

As the wells are brought together and separated adiabatically, the external states evolve according to Fig.~\ref{fig2}. That is, each vibrational eigenstate at $d\gg\sigma$ evolves continuously into its respective eigenstate at $d=0$. As $d$ decreases, the degeneracies between symmetric and antisymmetric eigenstates are lifted, resulting in a dynamic phase difference between $\frac{1}{\sqrt{2}}(|\widetilde{\psi^L\psi^R}\rangle+|\widetilde{\psi^R \psi^L}\rangle)$ and $\frac{1}{\sqrt{2}}(|\psi^L\psi^R\rangle-|\psi^R\psi^L\rangle)$, corresponding to the difference in respective energy curves (see energy curves 3e and 3f in Fig.~\ref{fig2}a, or 4c and 4d in Fig.~\ref{fig2}b). Furthermore, degeneracies between the even two-qubit states \{$|00\rangle$, $|11\rangle$, $|\Psi^+\rangle$\} are removed if the interaction strengths $a_{ij}$ differ, which is usually the case. This state-dependent interaction results in additional phase differences between qubit states of the same symmetry~\cite{Stock:2007}. Thus, each joint internal and external state acquires a unique phase, and the final state $|\psi_\text{f}\rangle$ upon re-separating the wells is,
\begin{gather}\label{finalstate}
\left|\right. \!\!\left.\psi_\text{f}\right> = (|\psi^L\psi^R\rangle-|\psi^R\psi^L\rangle)\otimes(\textstyle\frac{\alpha\nu-\beta\mu}{2}e^{-i\phi_-}\!\left|\Psi^-\right>) \\
\quad\quad\quad\quad + (|\widetilde{\psi^L\psi^R}\rangle+|\widetilde{\psi^R\psi^L}\rangle)\otimes(\textstyle\frac{\alpha\mu}{\sqrt{2}} e^{-i\phi_{00}}|00\rangle + \textstyle\frac{\alpha\nu+\beta\mu}{2}e^{-i\phi_+}|\Psi^+\rangle + \textstyle\frac{\beta\nu}{\sqrt{2}} e^{-i\phi_{11}}\left|11\right>) \,.\notag
\end{gather}
For positive scattering lengths, the phases are given by,
\begin{equation}
\phi_{jj} \equiv \frac{1}{\hbar}\int_{t_\text{i}}^{t_\text{f}}E^0_{|jj\rangle}\left(d(t)\right)\,\text{d}t\quad\text{and}\quad\phi_\pm \equiv \frac{1}{\hbar}\int_{t_\text{i}}^{t_\text{f}}E^{\frac{1}{2}\mp\frac{1}{2}}_{|\Psi^\pm\rangle}\left(d(t)\right)\,\text{d}t.
\end{equation}
Equation (\ref{finalstate}) is also valid for negative scattering lengths, although different phases will be acquired since $\frac{1}{\sqrt{2}}(|\widetilde{\psi^L\psi^R}\rangle\!+\!|\widetilde{\psi^R\psi^L}\rangle)$ is not the vibrational ground state when $a_{i\! j} < 0$.

Clearly, this evolution can be thought of as the identity acting on the vibrational subsystem tensored with a unitary $U$ acting on the qubit subsystem. Thus, we can discard the vibrational terms and examine the unitary evolution of the qubit subsystem by itself. Using matrix notation,
\begin{align}
\left|0\right>=
\begin{pmatrix}
1 \\ 0
\end{pmatrix} \qquad \text{and} \qquad
\left|1\right>=
\begin{pmatrix}
0 \\ 1
\end{pmatrix},
\end{align}
we can write $U$ as,
\begin{align}
U &=
\frac{1}{2}\begin{pmatrix}
2e^{-i\phi_{00}}&0&0&0 \\ 0& e^{-i\phi_+}\!+\!e^{-i\phi_-} &e^{-i\phi_+}\!-\!e^{-i\phi_-}&0 \\ 0&e^{-i\phi_+}\!-\!e^{-i\phi_-}&e^{-i\phi_+}\!+\!e^{-i\phi_-}&0 \\ 0&0&0&2e^{-i\phi_{11}}
\end{pmatrix} \\
 &= T
\begin{pmatrix}
e^{-i\phi_{00}}&0&0&0 \\ 0&e^{-i\phi_+}&0&0 \\ 0&0&e^{-i\phi_-}&0 \\ 0&0&0&e^{-i\phi_{11}}
\end{pmatrix}
T^\dag, \qquad \text{where} \qquad
T = \begin{pmatrix}
1&0&0&0 \\ 0&\frac{1}{\sqrt{2}}&\frac{1}{\sqrt{2}}&0 \\ 0&\frac{1}{\sqrt{2}}&-\frac{1}{\sqrt{2}}&0 \\ 0&0&0&1
\end{pmatrix}. \notag
\end{align}

This entangling operation is diagonal in the partial Bell basis $\{|00\rangle, |\Psi^\pm\rangle, |11\rangle\}$. As noted in\cite{Hayes:2007}, even if the interaction strengths are state-independent, the singlet state $\left|\Psi^-\right\rangle$ acquires a phase different from the triplet states (except in the limit as $a_{ij}\rightarrow\pm\infty$, when the gate is no longer feasible). State-dependent traps or atomic interactions generally result in state-dependent interactions with the environment as well, making the qubits more sensitive to noise. Avoiding this state-dependence leads to the inherent robustness observed in initial experiments~\cite{Anderlini:2007}, as compared to earlier experiments wherein gate fidelities were severely limited because of dephasing due to state-dependent traps~\cite{Mandel:2003}. Furthermore, this gate works for a wide range of positive scattering lengths, as we show elsewhere~\cite{Stock:2007}. This is especially important for experiments employing atomic species with unknown or approximately known scattering lengths (e.g., Yb or Sr).

A ``controlled phase'' gate (i.e., $e^{-i\pi|11\rangle\langle 11|}$) in the computational basis can be obtained by combining single qubit phase gates $S(\theta) = e^{-i\theta|1\rangle\langle 1|}$ with a pair of $U$ gates:

\begin{align}
G = U\left(S(\pi)\otimes S(0)\right)U
  = \begin{pmatrix}
     e^{-2i\phi_{00}}&0&0&0 \\ 0&e^{-i(\phi_++\phi_-)}&0&0 \\ 0&0&-e^{-i(\phi_++\phi_-)}&0 \\
     0&0&0&-e^{-2i\phi_{11}} \end{pmatrix}.
\end{align}

$G$ is locally equivalent to the ``tunable controlled-phase'' gate $e^{-i\gamma |11\rangle\!\langle11|}$, subject to the constraint:
\begin{equation}
\phi_{00} + \phi_{11} - \phi_+ - \phi_- = \left(2n\!\pm\!\textstyle\frac{1}{2}\right)\gamma, \quad\forall\;\, n \in \mathbb{Z}.
\end{equation}

While a simple controlled-phase gate is itself a universal entangling gate, many quantum algorithms (e.g., the quantum Fourier transform) can be performed more efficiently when tunable controlled-phase gates are available. Here, the value of $\gamma$ can be easily tuned, simply by adjusting the speed at which the optical tweezers are combined and separated.  The inherent robustness and easy tunability of this gate make it a highly desirable one for quantum information processing.

\section{Alternative entanglement preparation}

For some quantum information processing applications, universal entangling gates are not necessary and an ability to prepare entangled pairs will suffice. Atomic quantum repeaters based on entanglement swapping~\cite{Duan:2001} provide an example of one such application. We next examine a scheme which uses symmetrization requirements and a selective excitation to produce Bell pairs.

We begin with two bosonic atoms in the ground state of a single well, both with internal state $\left|0\right>$. The energy of this state is $E^0_{|00\rangle}(0)$. It is then possible to perform a coherent transition, selectively exciting to the eigenstate with energy $E^0_{|\Psi^+\rangle}(0)$. If the interaction for atomic qubits in the state $|\Psi^+\rangle$ is significantly different than that of $\left|11\right>$, one can deterministically excite only to the $|\Psi^+\rangle$ qubit eigenstate, since the state $\left|11\right>$ is off resonant and the overall state must remain symmetric. An excitation to the antisymmetric state $|\Psi^-\rangle$ is not possible as long as the two atoms are in the symmetric vibrational ground state. Thus, any initial population in antisymmetric vibrational states (e.g., due to heating) must be avoided to keep the fidelity of entanglement generation high. Furthermore, the vibrational spacing and sidebands due to the interaction energy must be spectroscopically resolvable. With typical vibrational energies on the order of kHz and on-site interaction shifts of close to 100Hz (dependent on atomic species but tunable by the tightness of traps), this selective excitation process is generally slow, but nevertheless viable. The final state after separating the atoms adiabatically is,
\begin{equation}
\left|\psi_\text{final}\right> = \textstyle\frac{1}{2}(|\widetilde{\psi^L\psi^R}\rangle\!
+\!|\widetilde{\psi^R\psi^L}\rangle)\otimes\left(\left|01\right>\!+\!\left|10\right>\right).
\end{equation}

This operation provides a novel way of creating Bell states deterministically, but does not constitute a universal two-qubit entangling gate. It does however allow for fundamental tests of quantum mechanics and Bell inequality violations, as well as basic quantum information processing and communication tasks.

\section{Speed constraints for adiabaticity}

\begin{figure}[h]
\begin{center}
$\begin{array}{c@{\hspace{1mm}}c}
\multicolumn{1}{l}{\mbox{\bf (a)}} &
    \multicolumn{1}{l}{\mbox{\bf (b)}} \\
\includegraphics[width=65mm]{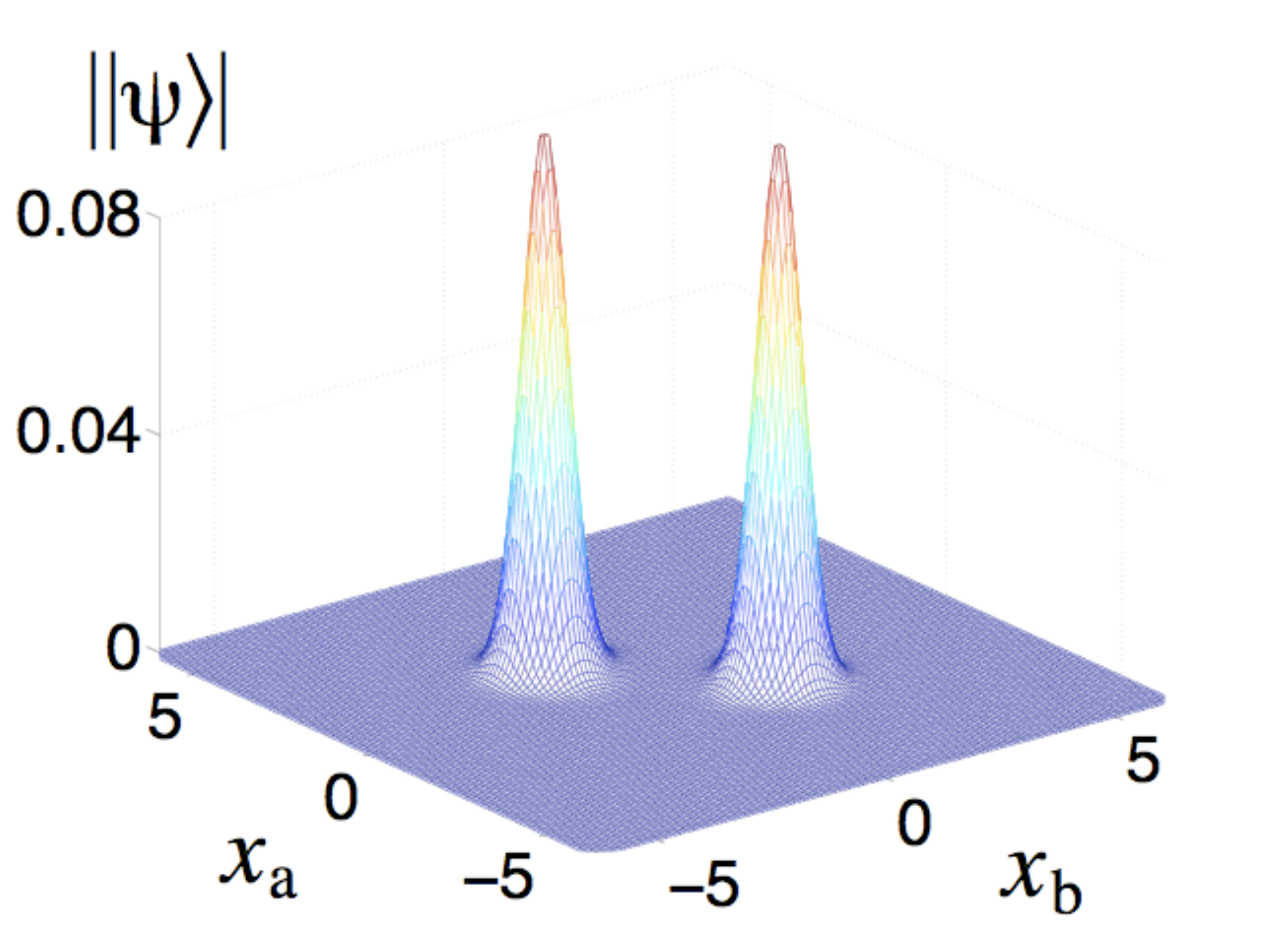} &
    \includegraphics[width=65mm]{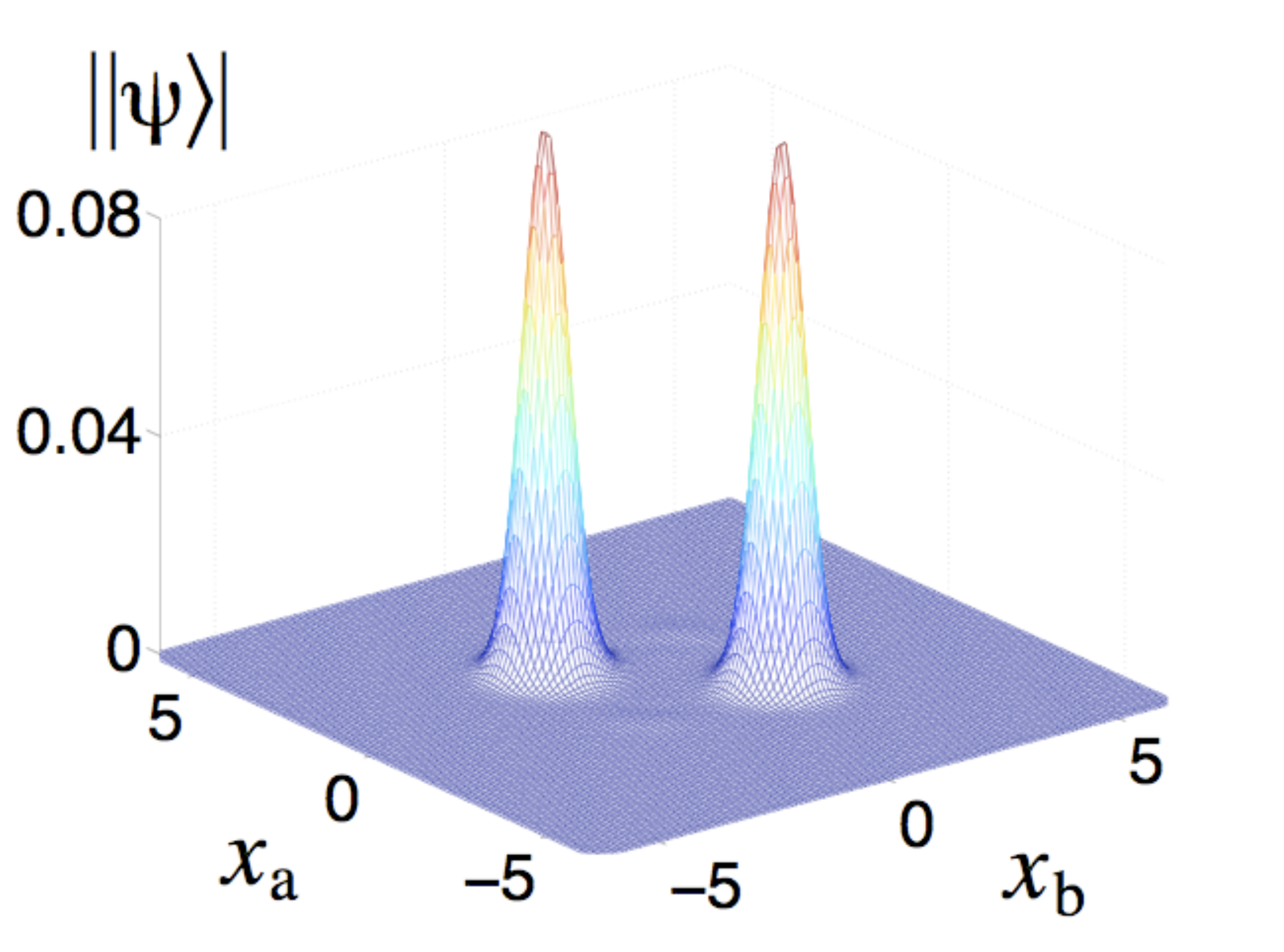} \\
\multicolumn{1}{l}{\mbox{\bf (c)}} &
    \multicolumn{1}{l}{\mbox{\bf (d)}} \\
\includegraphics[width=65mm]{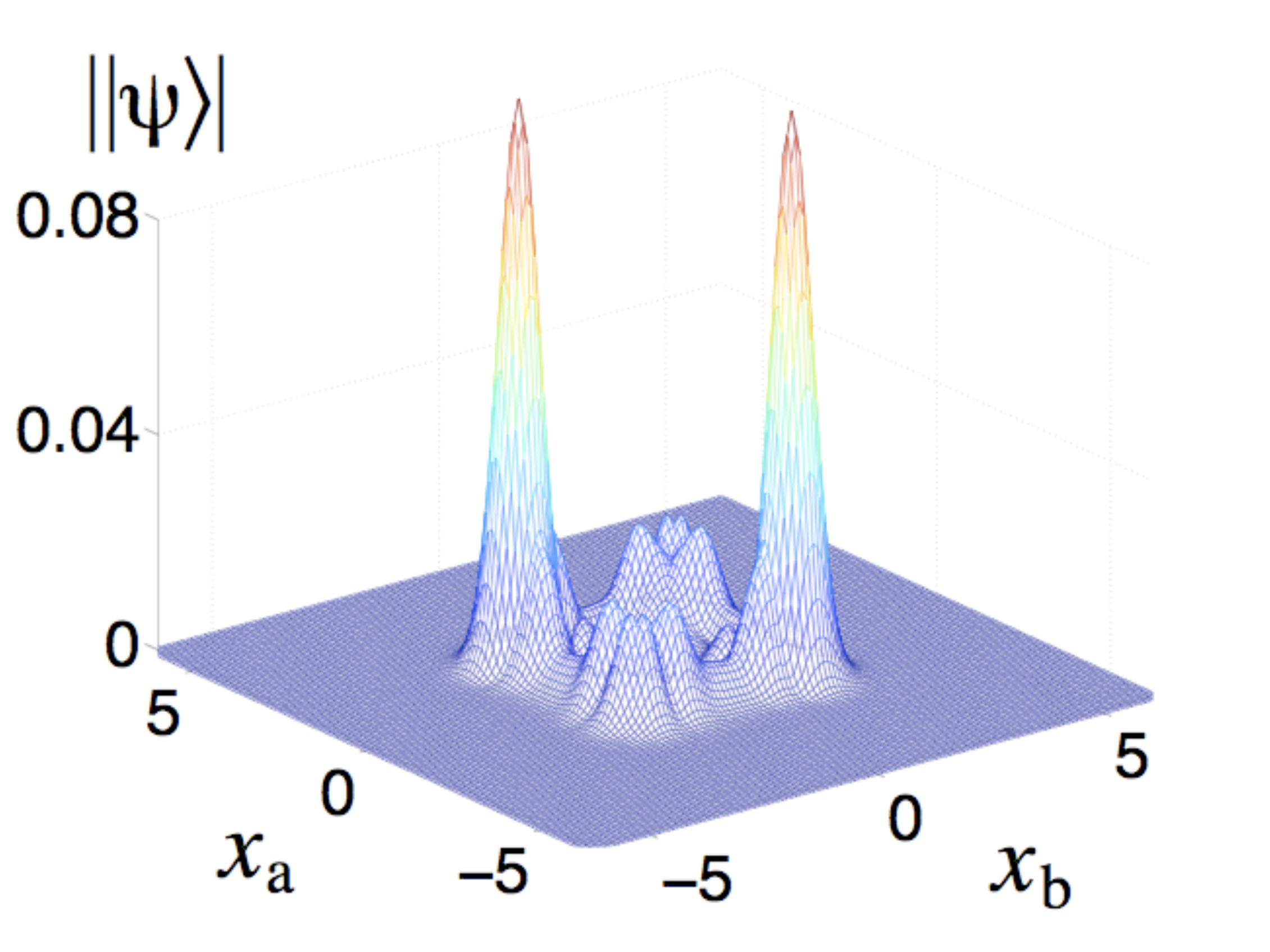} &
    \includegraphics[width=65mm]{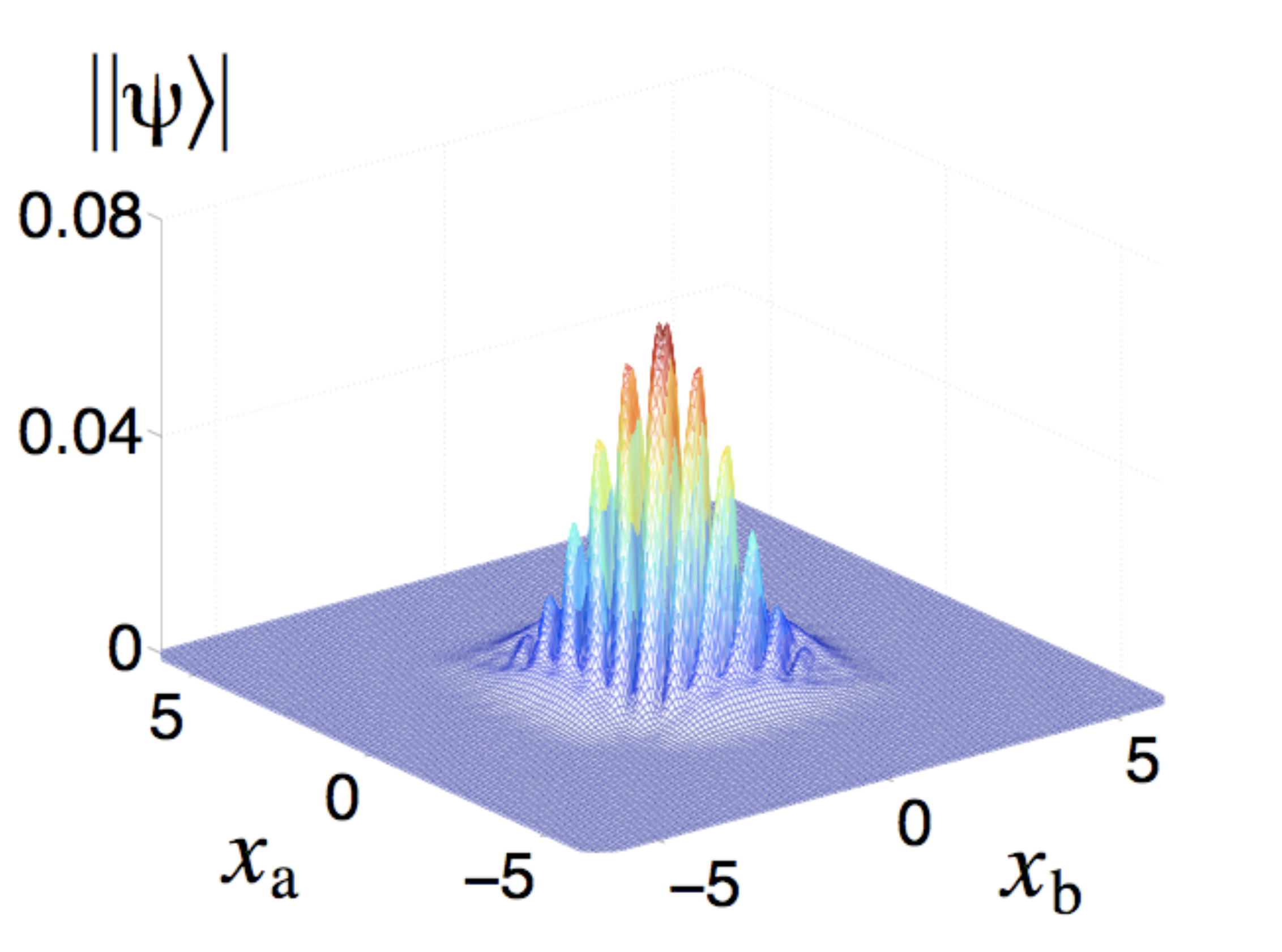}
\end{array}$
\end{center}
\caption{Snapshots of the magnitude of the two-atom vibrational wavefunction ($||\psi\rangle| \equiv ||\psi(x_\text{a},x_\text{b},t)\rangle|$) as a function of the position of each atom. Plot (a) shows the initial wavefunction, $|\psi_\text{init}\rangle = (|\widetilde{\psi^L\psi^R}\rangle+|\widetilde{\psi^R\psi^L}\rangle)/\sqrt{2}$. Plots (b-d) show the wavefunction after the wells have been brought together and separated. Initial conditions are the same for all figures, and only the well speed (in units of $v_o={\hbar\sigma\omega_{ab}^2}/{V_o}$) is varied. The resulting vibrational state fidelities $f = |\langle\psi_\text{init}|\psi\rangle|^2$ are as follows: (b) $v \approx 0.01v_o$, $f = 0.9997$. (c) $v \approx 0.1v_o$, $f = 0.491$. (d) $v \approx v_o$, $f = 0.002$.}
\label{fig3}
\end{figure}

An approximation to the general adiabaticity criterion is given in~\cite{Bransden}:
\begin{equation}
\left|\left<a\right|\textstyle\frac{\partial \hat{H}}{\partial t}\left|b\right>\right| \ll \hbar\omega_{ab}^2 \;\;\; \forall \; \left|a\right> \neq \left|b\right>,
\end{equation}
where $\omega_{ab} = \text{min}(\left|E_b(t)-E_a(t)\right|/\hbar)$ and where $\left|a\right>$ and $\left|b\right>$ are time-dependent eigenstates of an arbitrary Hamiltonian. Since our specific Hamiltonian is invariant under exchanges of both symmetry and parity, transitions between vibrational states of different symmetry or parity are suppressed. Thus, in our case $\omega_{ab}$ is determined by the energy gap of the two closest states having both equal symmetry and parity. This restriction contributes significantly to the robustness of this gate. Since only the double-well potential is time-dependent, the left side of the equation reduces to $|\frac{\partial V(x,d(t))}{\partial t}|$. Assuming constant $v$ and maximizing $|\frac{\partial V(x,d(t))}{\partial t}|$ with respect to $x$, we obtain the adiabaticity criterion:
\begin{equation}
v \ll {\hbar\sigma\omega_{ab}^2}/{V_o}.
\end{equation}

Time-dependent numerical simulations confirm the validity of this simple criterion over a wide range of values of $V_o$ and $a_{ij}$ (including $a_{ij}<0$). Plots of the two-particle wavefunction comparing adiabatic and non-adiabatic evolutions are shown in Fig. \ref{fig3}. Under adiabatic conditions, we recover both atoms in separate wells. Under non-adiabatic conditions both atoms may end up in the same well with non-negligible probability, resulting in an erroneous gate. However, time-dependent simulations have also shown significant revivals, with both atoms ending up in opposite wells with large probability even under non-adiabatic conditions. This suggests the very real possibility of producing a fast, coherent, non-adiabatic gate via optimal control.

\section{Conclusion}
In summary, we have proposed two schemes for preparing pairs of entangled atoms. We have shown it possible to construct a tunable universal entangling gate via the exchange interaction between identical bosons, promising high fidelity operation for positive (repulsive) and even negative (attractive) interaction strengths. This is of particular importance for quantum information processing applications that use novel species of atoms, such at Group II-like atoms (e.g., Yb and Sr)~\cite{Stock:2007}, for which the collisional interaction parameters are partially unknown. In addition, we have introduced a novel entanglement scheme allowing the creation of Bell pairs. This scheme could prove useful for quantum communication schemes and fundamental tests of quantum mechanics. The use of this entanglement operation for Group II-like atoms and its application to fundamental tests of quantum mechanics are studied in detail in other work~\cite{Stock:2007}.


\section{Acknowledgments}
We specially thank D. Hayes and I. Deutsch for insightful discussions on entangling atoms via exchange interactions. We further thank Michael Skotiniotis and Nathan Wiebe for helpful comments and discussion. This work was supported by NSERC, AIF, CIFAR, iCORE, MITACS, NSF, and The Welch Foundation.

\end{document}